\documentclass[pre,amsmath,amssymb,superscriptaddress,showpacs]{revtex4-2}
\usepackage{enumitem}
\usepackage{amsbsy}
\usepackage{latexsym}
\usepackage{color}
\usepackage{hyperref}
\usepackage{graphicx}
\usepackage{psfrag}
\usepackage[normalem]{ulem}
\usepackage{bm}
\usepackage{amsmath}
\usepackage{comment}
\usepackage{notes2bib}

\newcommand{\be}{\begin{equation}}
\newcommand{\ee}{\end{equation}}
\newcommand{\bea}{\begin{eqnarray}}
\newcommand{\eea}{\end{eqnarray}}

\begin{document}

\title{Ordering kinetics with long-range interactions: interpolating between voter and Ising models}

\author{Federico Corberi}
\email{fcorberi@unisa.it}
\affiliation{Dipartimento di Fisica ``E.~R. Caianiello'',
	via Giovanni Paolo II 132, 84084 Fisciano (SA), Italy.}
\affiliation{INFN, Gruppo Collegato di Salerno,
	via Giovanni Paolo II 132, 84084 Fisciano (SA), Italy.}
\author{Salvatore dello Russo}
\email{s.dellorusso@studenti.unisa.it}
\affiliation{Dipartimento di Fisica ``E.~R. Caianiello'',
via Giovanni Paolo II 132, 84084 Fisciano (SA), Italy.}
\author{Luca Smaldone}
\email{lsmaldone@unisa.it}
\affiliation{Dipartimento di Fisica ``E.~R. Caianiello'',
via Giovanni Paolo II 132, 84084 Fisciano (SA), Italy.}
\affiliation{INFN, Gruppo Collegato di Salerno,
	via Giovanni Paolo II 132, 84084 Fisciano (SA), Italy.}

\begin{abstract}
	We study the ordering kinetics of a generalization of the voter model with long-range interactions, 
	the $p$-voter model, in one dimension.  It is defined in terms of boolean variables $S_{i}$, agents or spins, 	
	located on sites $i$ of a lattice, each of which takes in an elementary move the state of the majority of $p$ 
	other agents at distances $r$ chosen with probability $P(r)\propto r^{-\alpha}$. For $p=2$ the model can be
	exactly mapped onto the case with $p=1$, which amounts to the voter model with long-range interactions decaying algebraically. For $3\le p<\infty$, instead, the dynamics falls into the universality class
	of the one-dimensional Ising model with long-ranged coupling constant $J(r)=P(r)$ quenched to 
	small finite temperatures. In the limit $p\to \infty$, a crossover to the (different) behavior of the 
	long-range Ising model quenched to zero temperature is observed.  Since for $ p > 3$ a closed set of differential equations cannot be found, we employed numerical simulations to address this case.
\end{abstract}
\maketitle

\section{Introduction} \label{SecIntro}

Studies on the structure of  matter are very often approached by means of effective theories where the complexity of the original problem is somehow reduced. Thermodynamics and hydrodynamics, just to quote a couple of examples, are effective descriptions in terms of few variables (like pressure, volume, viscosity etc...) of a by far more complex microscopic reality where Avogadro numbers of more
elementary constituents interact among themselves and, possibly, with the environment. Such effective theories provide prescriptions to compute at a quantitative level the quantities entailed. 
Reduction of degrees of freedom results in a poorer resolution, because only the regularities of a small set of variables emerging
from the collective behavior can be considered. However, this does not mean that predictions, for such entailed quantities, are necessarily less precise, as the previous examples show. On the other hand, effective theories have the great advantage of being
sometimes much more manageable than more fitting descriptions, making their use so widespread and almost unescapable. The question of how close
effective theories are to the underlying reality, however, is generally open and 
must be addressed case by case. 

Perhaps one of the most studied, simple and paradigmatic systems with many interacting agents are
classical spin models, like the Ising one. One of their merits has been to allow a deep comprehension
of the basic features of phase-transitions.
Despite their simplicity, they provide a correct description of these phenomena, at least for
some observable, due to the universality concept. Namely, despite operating a drastic simplification of reality,
these models are capable to retain the few relevant ingredients whereby the collective behavior is formed. The many-body character of these system is even
more pronounced when the interaction, instead of being among nearest neighbors, as  in the simplest modelizations, is all to all by means of a 
long-range coupling. 

In this paper we study a model with long-range interactions,
the $p$-voter model (pVM), where spins $S_i=\pm 1$, hosted on the sites $i$ of a lattice, align with the direction of the majority of $p$ other spins chosen randomly at distance $r$ from $i$ with probability $P(r)$ with a fat tail,
typically decaying as $P(r)\propto r^{-\alpha}$. 
A similar model, however of a mean-field character (corresponding to $\alpha=0$), was introduced and studied in~\cite{PhysRevE.92.052105,Chmiel2018}.
The pVM interpolates 
between the usual voter model (VM)~\cite{PhysRevE.83.011110,CorbCast24,corsmal2023ordering,corberi2024aging,corberi2024coarseningmetastabilitylongrangevoter} (with similar long-range interactions), corresponding to $p=1$, and the Ising model (IM) with 
coupling constant $J(r)=P(r)$, for $p\to \infty$. Therefore, one could argue that, for a generic finite value of $p$, the pVM could embody a reliable effective theory for the IM, at least for sufficiently large $p$. Needless to say, 
having a $p$-body interaction, instead of the all-to-all one of the IM, represents
a simplification at least from the point of view of numerical simulations. On the
analytical side, neither the IM nor the pVM with $p>2$ are solvable, however
let us mention that for the cases with $p=1,2$ analytical solutions of the pVM are at hand. The relations between the IM and the pVM, restricting to $p=3$,
were studied, in stationary states, in~\cite{PhysRevB.107.224204}.
In this article we focus on the
phase-ordering kinetics~\cite{Bray94,Corberi2010GrowingLS} of the system, starting from a fully disordered state,
on a one-dimensional lattice, for generic values of $p$. 
In particular, since the the two limiting cases, the VM and the IM, are known to behave differently~\cite{CLP_review,CLP_epl}, we answer the questions of how the pVM interpolates between them, how good the effective description provided by the pVM is and how large the value of $p$ must be in order to near the IM within a given tolerance.

Our study is not only a contribution to the understanding of the ordering kinetics of long-range systems~\cite{Bray94,CLP_review,CMJ19,iannone2021, Corberi_2019, CLP_epl, PhysRevE.103.012108, BrayRut94, RutBray94, CMJ19}, but has also a broader scope covering the
more general issue of effective descriptions in various fields of knowledge, an example of which will be discussed in Sec.~\ref{SecConcl}.

The phase-ordering kinetics considered in this paper amounts to the formation and growth of ordered domains of the equilibrium phases. This is accompanied by the reduction of the interfaces density $\rho(t)$, namely the amount of anti-aligned spins (per lattice site) 
in the system.
The analysis of the pVM conducted in this paper is based on this quantity, which usually behaves as 
\be
\rho(t)\propto t^{-a},
\label{eqRho}
\ee
$a$ being an universal exponent whose value changes among different universality classes. For the IM, $a$ is different for a quench
to a finite but small temperature, which we indicate as $T\to 0$, and for 
$T\equiv 0$. {\it Small temperature} means here a temperature small enough to have a long-lasting coarsening stage, before eventual equilibration.
The main results of our study is the existence of a value $p^*=3$ of $p$,
such that, by looking at the value of $a$, for $p^*\le p<\infty$ the pVM can be classified as being in the same universality class of the IM quenched to a finite temperature $T\to 0$, while for $p=2$ the model can be exactly mapped onto the VM. 
Letting $p\to \infty$, instead, the pVM crosses over to the universality
class of the IM quenched right at $T\equiv 0$. The difference between the  the cases with finite $p\ge p^{*}$ and 
infinite $p$ is due to the randomness in the choice of the $p$ neighbors,
which effectively mimics a temperature, which is suppressed as 
$p\to \infty$.

This paper extends the results obtained for the VM in a series of companion papers~\cite{CorbCast24,corsmal2023ordering,corberi2024aging,corberi2024coarseningmetastabilitylongrangevoter}, where many details can be found, which are useful in the case of the pVM. This set of articles, in turn, generalize the analytical studies~\cite{Frachebourg1996,Ben1996}
on the voter model to the case with long-range interactions. This represents a natural setting where the important issue of the universality properties, as $p$ is varied, can be studied in a clear and controlled way. We believe that this kind of study may help understanding the ordering kinetics of ferromagnetic models in a general perspective. 

The paper is organized as follows: 
In Sec.~\ref{SecModels} we introduce the models considered in this 
study, prove the equivalence between the pVM with $p=1$ and $p=2$,
and discuss the expected convergence of the pVM towards the IM
as $p\to \infty$.
In Sec.~\ref{SecRecalls} we review the known properties of the 
VM and of the IM.
The following Sec.~\ref{SecSimul} contains a discussion of our 
numerical results on the kinetics of the pVM for different values of $p$.
Finally, in Sec.~\ref{SecConcl} we discuss further some results
and some open points for further investigations, and we conclude the Article.

\section{Models} \label{SecModels}

The voter model (VM) considered in this paper is defined in terms of boolean (or spin) variables $S_i=\pm 1$, where $i=1,\dots,N$
runs over the sites of a one-dimensional lattice with periodic boundary conditions. In an elementary move, a randomly chosen variable $S_i$ takes the state of another, $S_k$, 
selected at distance $r$ with probability
\be
P(r)=\frac{1}{2Z}\,r^{-\alpha},
\label{eqP}
\ee 
where $Z=\sum_{r=1}^{N/2}r^{-\alpha}$.
The limit $\alpha \to \infty$ amounts to interactions limited to nearest neighbors, which is the original version of the model 
firstly introduced to study genetic correlations \cite{Kimura1964,1970mathematics}. Later on, its basic properties were derived in Refs. \cite{Clifford1973,Holley1975} and widely studied along the years \cite{Clifford1973,Holley1975,liggett2004interacting,Theodore1986,Scheucher1988,PhysRevA.45.1067,Frachebourg1996,Ben1996,PhysRevLett.94.178701,PhysRevE.77.041121,Castellano09}, with application to various disciplines \cite{Zillio2005,Antal2006,Ghaffari2012,CARIDI2013216,Gastner_2018,Castellano09,Redner19}.
Since than, several modifications of the original model have been proposed, to make it more suited to different situations \cite{Mobilia2003,Vazquez_2004,MobiliaG2005,Dall'Asta_2007,Mobilia_2007,Stark2008,Castellano_2009,Moretti2013,Caccioli_2013,HSU2014371,PhysRevE.97.012310,Gastner_2018,Baron2022}. 

With the kinetic rule described above, detailed balance is not obeyed, except for nearest neighbors interactions, where the VM can be mapped exactly onto the IM (also with nearest neighbors). Notice that, once $k$ has been chosen, the 
new state of spin $S_i$ is found deterministically. In this sense, the algorithm may naively resemble
a zero-temperature process. However, since the choice of $k$ is stochastic,  
$S_i$ can flip in a single update with a certain probability that can be written~\cite{CorbCast24} as
\be
W_{VM}(S_i)=\frac{1}{2N}\sum _{r=1}^{N/2}P(r)\sum _{k=[[i\pm r]]}(1-S_iS_k),
\label{transrates}
\ee
where the factor $1/N$ describes the probability to randomly pick up $S_i$.
The symbol $[[\dots]]$
means that the boundary conditions must
be correctly taken into account, namely
\be
[[q]]=\left \{
\begin{array}{lcl}
q, & \quad \mbox{if} &\quad 1\le q\le N, \\
q-N, &\quad  \mbox{if} &\quad q>N \\
q+N, &\quad \mbox{if} &\quad q<1,
\end{array}
\right .
\ee
for $-N+1\le q \le 2N$, $q\in \mathbb{N}$. This model is exactly solvable and its dynamics has been studied numerically in~\cite{PhysRevE.83.011110} and analytically in~\cite{CorbCast24}, for all values of $\alpha$.

We introduce now the $p$-voter model (pVM), a generalization of the above VM, where detailed balance is still violated (this will be explicitly shown later on), and $S_i$ takes the orientation of the partial spin-average 
\be
H_i(\{S_{k_n}\}_1^p)=\frac{1}{p}\,{\sum_{n=1}^p}^* S_{k_n}
\label{eqH}
\ee
of $p$ spins $S_{k_n}$, each chosen at different distances $r_n=|k_n-i|$ 
(here and in the following, all distances are computed with respect to site $i$, so, in order to lighten the
notation, we drop the $i$-dependence and write simply $r_n$ instead of, say, $r_{i,k_n}$) with probability $P(r_n)$.
Clearly, being chosen randomly, a spin can occur more than once in the average, particularly for large $p\gtrsim N$. 
The asterisk attached to the sum indicates that $S_i$ must be excluded, namely $k_n\neq i$, $\forall n=1,p$.
The argument $\{S_{k_n}\}_1^p$
of this function reminds us that it depends on all the spins $S_{k_n}$, with $n=1,p$. From now on, to make
the notation more evident, we reserve the 
letters $i,j,k$ to denote lattice sites; other letters will be used for different purposes. In the mean-field limit $\alpha \to 0$ only, this model, dubbed as $q$-Ising, was originally introduced in~\cite{PhysRevE.92.052105,Chmiel2018},
where the stationary states were studied finding that a phase-transition 
can only occur for $p>2$, and is of the continuous type for $p=3$
and discontinuous for $p>3$.
The transition probability of the pVM is readily obtained upon generalizing expression~(\ref{transrates}), and reads
\be
W_{pVM}(S_i)=\frac{1}{2N}\prod _{m=1}^p \,\sum _{r_m=1}^{N/2}P(r_m)\sum _{k_m=[[i\pm r_m]]}[1-S_i \sigma (H_i)],
\label{transp}
\ee
where 
\be
\sigma (H)=
\left \{
\begin{array}{lcr}
\mbox{sign}(H) &, & \mbox{if }H\ne 0, \\
0 &, & \mbox{if } H=0,
\end{array} \right .
\label{eqSigma}
\ee
is the majority rule.
Clearly, the original VM is recovered for $p=1$. The 2VM, namely the model 
with $p=2$ can also be mapped onto the VM. Indeed, in this case, the
transition probability amounts to  
\be
W_{2VM}(S_i)=\frac{1}{2N}\sum _{r_1,r_2}^{1,N/2}P(r_1)P(r_2)\sum _{k_1=[[i\pm r_1]]}\sum _{k_2=[[i\pm r_2]]}[1-S_i H_i(S_{k_1},S_{k_2})],
\label{trans2}
\ee
because, for $p=2$, $H_i$ runs only over the three integers $(-1,0,+1)$ (see definition~(\ref{eqH})). Using the expression~(\ref{eqH}) one has
\begin{eqnarray}
&&W_{2VM}(S_i)=\frac{1}{2N}\sum _{r_1,r_2}^{1,N/2}P(r_1)P(r_2)\sum _{k_1=[[i\pm r_1]]}\sum _{k_2=[[i\pm r_2]]}\left[1-\frac{1}{2}S_i (S_{k_1}+S_{k_2})\right]=\nonumber  \\
&=&
\frac{1}{2N}\left \{ \frac{1}{2}\left [\sum _{r_2=1}^{N/2}P(r_2)\sum _{k_2=[[i\pm r_2]]} 1\right ]\sum _{r_1=1}^{N/2}P(r_1)\sum _{k_1=[[i\pm r_1]]}\left(1-S_i S_{k_1}\right )+ (1\leftrightarrow 2) \right \}= \nonumber \\
&=& \frac{1}{2N}\sum _{r=1}^{N/2}P(r)\sum _{k=[[i\pm r]]}\left(1-S_i S_k\right),
\label{trans22}
\end{eqnarray}
where $(1\leftrightarrow 2)$ means the first term in curly brackets with the exchange of the labels $1$ and $2$,
and we have used $\sum _{r=1}^{N/2}P(r)\sum _{k=[[i\pm r]]} 1=1$, due to the normalization of $P$
(see Eq.~(\ref{eqP})). Comparing this form with the one~(\ref{transrates}) of the VM, one concludes that 
the 2VM can be exactly mapped onto the VM. This is confirmed with great precision by running numerical simulations of the two models. However, such equivalence is limited to $p=2$ where $\sigma$ is linear in $H$, which is not the case when $p>2$. 
Indeed, it is easy to prove with a similar calculation (see Appendix~\ref{Sec_App}) that 
a modified pVM where $\sigma (H)$ is 
replaced by the linear function $\sigma (H)=H$ is equivalent to the VM for any $p$. $p=2$ is the only case where $\sigma (H)$ is automatically linear, as we have thus shown. This may already suggest that the model with $p\ge 3$ might behave in a radically different way from the VM.
Before moving to the next topic, let us remark that the model introduced insofar is different
from the one introduced and studied in~\cite{Castellano_2009,Moretti2013,PhysRevE.86.011105}, there denoted as $q$-voter model.

The third model we consider in this Article is the one-dimensional Ising model (IM) with long-range
algebraic interactions~\cite{PhysRevLett.29.917}, whose Hamiltonian is
\be
{\cal H}(\{S_i\})=-{\sum _{i,j=1}^{N}}^* P(r_{ij})S_iS_j,
\label{Ham}
\ee
with $r_{ij}=|i-j|$ and $P(r_{ij})$, defined as in Eq.~(\ref{eqP}), is here a coupling constant (not a probability).
The asterisk close to the sum means that $j\ne i$.
Notice that we use symbols as to reveal the similarities with the voter models introduced before,
therefore we prefer to write $P$ for the interaction constant, even if it may look unusual.
$Z$, in this context, is
the Kac factor which guarantees an extensive energy also for $\alpha \le 1$.
The Glauber flip probability, in a process at zero temperature, is
\be
W_{IM}(S_i)=\frac{1}{2N}[1-S_i \sigma(\hat H_i)],
\label{transIsing}
\ee
where 
\be
\hat H_i\left (\{S_k\}_1^{N-1}\right )={\sum _{j=1}^N}^* P(r_{ij})S_{j}=\sum_{r=1}^{N/2} P(r) \sum _{k=[[i\pm r]]}S_k
\label{eqHatH}
\ee
is the Weiss local field.
This quantity differs from $H_i$ in Eq.~(\ref{eqH}) because all the spins (except $S_i$) in the
system are deterministically summed up in the average, each counted once. Let us notice that the
transition probability~(\ref{transIsing}) can be written, more generally, for processes at finite temperature $T$, as
\be
W_{IM}(S_i)=\frac{1}{2N}[1-S_i \tanh (\beta \hat H_i)],
\label{transT}
\ee
with $\beta=1/(k_BT)$, $k_B$ being the Boltzmann constant, which clearly reduces to~(\ref{transIsing})
for $\beta \to \infty$. 
The flipping probabilities~(\ref{transIsing},\ref{transT}) obey detailed balance with respect to the Hamiltonian~(\ref{Ham}).
The kinetic of the Ising model with algebraic interactions cannot be exactly solved. However it has been  
studied numerically and with approximate analytic approaches in~\cite{CLP_review,iannone2021,Corberi_2019,CLP_epl,BrayRut94,RutBray94}.

The pVM can be regarded as a modification of the IM where, instead of $P(r_{ij})$ in Eq.~(\ref{eqH}) there are boolean coupling constants $J_{ij}=0,1$, which are dynamic variables taking the value $J_{ij}=1$ with a probability $P(r_{ij})$, and  with a further constraint $\sum _j J_{ij}=p$. This immediately shows that detailed balance is not obeyed. Indeed, in order to have detailed balance, the evolution of the $J_{ij}$ should be triggered by the Hamiltonian~(\ref{eqH}), therefore by the actual value of the spins, whereas this does not happen in the pVM.

In view of the comparison between the IM and the pVM, it is useful to consider 
the behavior of the latter for $p\to \infty$, which we denote as the $\infty$VM. Notice that $\infty VM$ means that the limit $\lim _{p\to \infty}$ is taken and not that 
$p$ is strictly set to infinity, which, by the way, would be impossible to do, at least numerically. This specification will be important in the following.

Writing the average $H_i$ as
\be
 H_i(\{S_{k_n}\}_1^p)={\sum_{j=1}^N}^* f_j S_j,
\ee
 where $j$ runs over the lattice sites and $f_j$ is the frequency of extraction of site $j$, i.e.
 $N_j/p$, where $N_j$ counts the number of times $S_j$ is chosen. 
In the large-$p$ limit, using the law of large numbers, 
this can be expressed as
\be
\lim_{p\to \infty} H_i(\{S_{k_n}\}_1^p)={\sum_{j=1}^N}^* P(r_{ij}) S_j=
\hat H_i\left (\{S_k\}_1^{N-1}\right ),
\label{frequencies}
 \ee
hence the average $H_i$ of the $\infty$VM equals the local field $\hat H_i$ of the IM.
After plugging the above large-$p$ expression for $H_i$ into Eq.~(\ref{transp}),
recalling that 
$\sum _{r_m=1}^{N/2}P(r_m)\sum _{k_m=[[i\pm r_m]]}=1$,
because of the chosen normalization of $P$ (Eq.~(\ref{eqP})), the transition probability~(\ref{transp})
of the $\infty$VM equals the one~(\ref{transIsing}) of the IM at $T=0$. A word of caution, however, must be added.
The argument above maps the $\infty$VM into the IM at $T=0$ after neglecting the fluctuations of the multiplicities $N_{k_m}$. However, for large but yet finite $p$ some fluctuations will be present, although small. Then the question
arises: are the two models really equivalent? The answer to this question depends on the stability of the behavior
of the IM at $T=0$ with respect to the introduction of small disturbances, mimicking the tiny fluctuations of the 
$N_{k_m}$, which we can imagine to be akin to a small temperature. Then we can argue that, if the IM is stable
with respect to such disturbances, then one can conclude that the $\infty$VM can be mapped onto the IM at $T=0$,
otherwise such mapping is not expected necessarily to hold and, possibly, one could find equivalence between 
the $\infty$VM and a {\it disturbed} IM, possibly the IM at finite (but small) $T$. If the process considered is not
stationary, such as in ordering kinetics, the kind of equivalence observed may depend on the observation time.
As a final remark, let us stress that, in the previous considerations, the limit $p\to \infty$ must be taken with
fixed $N$ and, therefore, the thermodynamic limit must implemented later. Indeed, one can argue that the 
$\lim _{p\to \infty}$ amounts to $p\gg N$.

This whole matter will be investigated and further discussed in Sec.~\ref{SecSimul}.    
 
\section{Behavior of the VM and IM: a reminder} \label{SecRecalls}

In this section we recall the kinetic properties of the VM and of the IM, evolving from an initial disordered state
with $\langle S_i\rangle=0$ and $\langle S_iS_j\rangle=\delta_{ij}$. Here and in the rest of this article we will mainly use
the density of interfaces $\rho(t)$, i.e. the total number of misaligned spins divided by $N$, as a reference observable to frame the kinetic behavior.
The decay of quantity, which is a natural representation of the degree of ordering of the system and usually occurs
 in the form~(\ref{eqRho}),
provides information on the dynamical exponent $a$, which is expected to be universal, i.e. to be able to discriminate among different universality classes. A further advantage in studying $\rho$ is
its relatively easy computation in numerical simulations. 
In order to validate further our conclusions, we will also 
compute another observable, the autocorrelation function, whose 
properties in the various models will be discussed further on at
the end of Sec.~\ref{a>1}.

The behavior of the VM and of the IM is different for different ranges of $\alpha$-values
$0\le \alpha _{LR} <\alpha _{SR}$, where $\alpha_{LR}=2$ and $1$ for the VM~\cite{CorbCast24} and the IM, respectively,
and $\alpha_{SR}=3$ and $2$, again for the VM~\cite{CorbCast24} and IM~\cite{CLP_review, BrayRut94, RutBray94}. 
$\alpha _{SR}$ is a critical value above which asymptotically $a=1/2$,  
as in the nearest neighbor case. $\alpha _{LR}$, instead, is the value of $\alpha$ below which a mean-field phenomenology, in a broad sense, starts to be observed.
By this, we mean that a true coarsening behavior, akin to the one observed for short-range interactions, is
not displayed. In the voter model  the mean-field phenomenology amounts to the presence of partly correlated metastable stationary states, becoming infinitely long-lived and preventing complete ordering in the thermodynamic limit $N\to \infty$. This means that $\rho$ remains constant, as expressed by $a=0$ in the first line and in the rightmost entry of the second line of table~\ref{table_a}. Notice that for
$1<\alpha \le \alpha _{LR}$ there is a pre-asymptotic coarsening regime characterized by $a=1$, as indicated in the leftmost part of the second line of the table. This means that, before entering in the stationary state, the system experiences a coarsening stage, whose duration diverges in the thermodynamic limit, characterized by that value of the exponent. 

In the Ising model there are no such states, and the mean-field 
phenomenology is manifested by the fact that, for $\alpha \le \alpha_{LR}=1$, the statistical ensemble is dominated by histories where ordering is 
achieved by aligning all the spins with the sign of the majority in a time of order one, without formation of domains, as it happens in the fully mean-field case
($\alpha =0$). Intuition that $\alpha_{LR}$ is the value $\alpha=1$ where the interaction becomes no-longer integrable is correct, whereas for the VM, where detailed balance is violated, there is the different value
$\alpha_{LR}=2$.  Note that $\alpha_{SR}$ and $\alpha_{LR}$ are larger by one unit for the VM with respect to the Ising one, which can be understood because in the latter the interaction is mediated by the Weiss field $\hat H$, which is a quantity formed by the contribution of all the spins, while in the VM the interaction is one to one. Since the Weiss field implies a summation over all the system, which is not present in the VM, simple power counting 
suggests that the critical dimensions 
are shifted by one, as indeed it is found.

Concerning the dynamic exponent, its asymptotic value is reported, in the various $\alpha$-ranges, in table~\ref{table_a}. For the IM, the behavior observed after a quench to sufficiently low but finite temperature is reported (the case of a quench to $T=0$ will be discussed further on). In this case some pre-asymptotic regimes, characterized by algebraic decay of $\rho$ with
$a$-values different from the asymptotic one, are observed for $\alpha >\alpha_{LR}$~\cite{CLP_review,CLP_epl}, which are listed in 
table~\ref{table_a} in different columns (leftmost corresponds to earlier regimes). Notice the fact that
the regime with $a=1/\alpha$ is truly asymptotic for $\alpha _{LR}<\alpha \le \alpha_{SR}$, where an ordered phase at low temperatures exists~\cite{Dyson1969}, whereas it is only 
transient for $\alpha > \alpha _{SR}$. 
Indeed, for $\alpha>\alpha_{LR}$ there is no phase-transition and, for this reason, a finite-temperature quench necessarily ends, after a coarsening stage with the various exponent reported in the table, into an equilibrium 
state with constant $\rho$. This explains the zero in the last column.
It is very important to notice that all the pre-asymptotic 
regimes indicated for the IM extend their duration indefinitely as $T\to 0$. 
Therefore, if $T=0$ is strictly 
considered, for $\alpha >\alpha_{LR}$ the exponent $a=1$ is observed at any time. Then, recalling the discussion 
at the end of the previous section, we must arrive to the conclusion that the asymptotic value of the exponent $a$ is not robust against thermal fluctuations (whatever small). Indeed it changes its value from $1$ to $1/\alpha$ or to 
$1/2$, depending on $\alpha_{LR}<\alpha \le \alpha _{SR}$ or $\alpha >\alpha_{SR}$, as soon as an infinitesimal temperature is switched on.

A different and more complex situation occurs in the IM for $\alpha \le \alpha_{LR}$. In this case it was shown~\cite{iannone2021}
that, after a quench to $T=0$, the statistical ensemble splits into two kind of evolutions. A fraction $1-P_\alpha$ of the trajectories
has a mean-field-like behavior, in that in a microscopic time all spins align with the average magnetization
seeded by the initial state (such magnetization, due to the central limit theorem, is of order $N^{-1/2}$ (per spin)),
so that there is no formation of domains, hence no coarsening, and the system very quickly approaches the
magnetized state. However, there exist also a fraction $P_\alpha$ of kinetic histories where domains form,
and coarsening is observed. Focusing only on this subset of trajectories, in~\cite{CORBERI2023113681}, it was conjectured that $a=1$. However, a clear-cut indication of this is lacking. This is represented by the question mark present in the line
$0\le \alpha \le \alpha_{LR}$, for the IM. On the other hand, the behavior of the same model, still for $\alpha \le \alpha_{LR}$, after a quench to a finite temperature, has never been investigated.

Moving again to the VM, we mentioned already a pre-asymptotic coarsening stage with $a=1$, before reaching stationarity, 
for $1<\alpha \le \alpha _{LR}$ (left of second line). Coarsening is truly asymptotic
with $a=1/(\alpha -1)$ or $a=1/2$ for $\alpha _{LR}<\alpha \le \alpha _{SR}$ or $\alpha >\alpha _{SR}$, respectively (third and fourth line of table~\ref{table_a}).

\begin{table}[ht]
	\centering
\begin{tabular}{||p{3.5cm}||p{3cm}|p{3cm}|p{3cm}|p{3cm}|}
	\hline \hline
	\multicolumn{5}{|c|}{\hspace{-1.5cm}\bf{VM}} \\
	\hline \hline
	$0\le \alpha \le 1\, $ &
	\multicolumn{4}{|c|}{0} \\
	\hline
	$1< \alpha \le \alpha_{LR}\, (=2)$ &\multicolumn{4}{|c|}{0} \\
	\hline
	$\alpha_{LR}<\alpha\le\alpha_{SR}\,(=3)$ &\multicolumn{4}{|c|}{$(\alpha-2)/(\alpha -1)$}\\
	\hline
	$\alpha >\alpha_{SR}$ & \multicolumn{4}{|c|}{$1/2$}\\ 
	\hline \hline
		\multicolumn{5}{|c|}{\hspace{1cm} {\bf IM} \hspace{1cm} (time $\to$)} \\
	\hline \hline
	$0\le \alpha \le \alpha_{LR}\,(=1)$ &\multicolumn{4}{|c|}{1?} \\
	\hline
	$\alpha_{LR}<\alpha\le \alpha_{SR}\,(=2)$ & \hspace{1.5cm}1 & \multicolumn{3}{|c|}{$1/\alpha$} \\
	\hline 
	$\alpha >\alpha_{SR}$ &\hspace{1.4cm} 1 &\hspace{1.3cm} $1/\alpha$ &\hspace{1.3cm}1/2 & \hspace{1.3cm} 0\\
	\hline \hline
\end{tabular}
\caption{Value of the exponent $a$ for (top of the table) the VM (after Ref.~\cite{CorbCast24}) and the for (bottom of the table) the IM quenched to a sufficiently small but finite temperature (after Ref.~\cite{CORBERI2023113681} (first line, $0\le \alpha \le \alpha _{LR}$) and Ref.~\cite{CLP_review} (second and third line, $\alpha >\alpha _{LR}$)), for different ranges of $\alpha$, specified in the left column (values of $\alpha_{LR}$ and $\alpha_{SR}$ are also indicated in parenthesis). In some cases there are different temporal regimes, corresponding to different $a$ values, separated in columns (time is increasing from left to right). 
The entry $1?$ for $0\le \alpha \le \alpha_{LR}$ indicates the putative nature
of this value, see text for further discussion.}
\label{table_a}
\end{table}
 
\section{\lowercase{p}VM: Numerical simulations} \label{SecSimul}

According to the discussion at the end of Sec.~\ref{SecModels}, for sufficiently large values of $p$, the behavior of the pVM is expected to belong to the same universality class of the IM. Referring to table~\ref{table_a}, the difference between the VM and the IM universality classes can be easily spotted by looking at asymptotic value of the exponent $a$ (at least for $\alpha <3$). A natural question is, therefore, how
the system crosses-over from the behavior of the cases with $p=1,2$, in the voter universality class, to the large-$p$ one of the Ising type. And, related to that, one would like to understand which is the characteristic value $p^*>2$ around which the cross-over occurs, and how sharp it is. We have investigated this issues by means of
numerical simulations of the pVM. For $p=1,2$ we reproduce with good accuracy the analytical results reported in table~\ref{table_a} (not shown here). In the following we discuss our results for the values $p>2$, for which the model is not analytically solvable. All the data presented are restricted
to times when finite-size effects 
are absent or, in some cases, 
are only incipient (this will be discussed later in specific cases).
Also, the system size is sufficient to avoid any kind of equilibration or stationarization. 
Each numerical run realizes a non-equilibrium statistical average, namely over different initial conditions and random time-trajectories (up to $5\times10^4$ in some cases).
We start with the case $\alpha >1$, where the behavior of the IM, to which we want to compare, is better understood. 

\subsection{\boldmath{$\alpha >1$}} \label{a>1}

\begin{figure}[ht]
	\vspace{1.0cm}
	\centering
	\rotatebox{0}{\resizebox{1.0\textwidth}{!}{\includegraphics{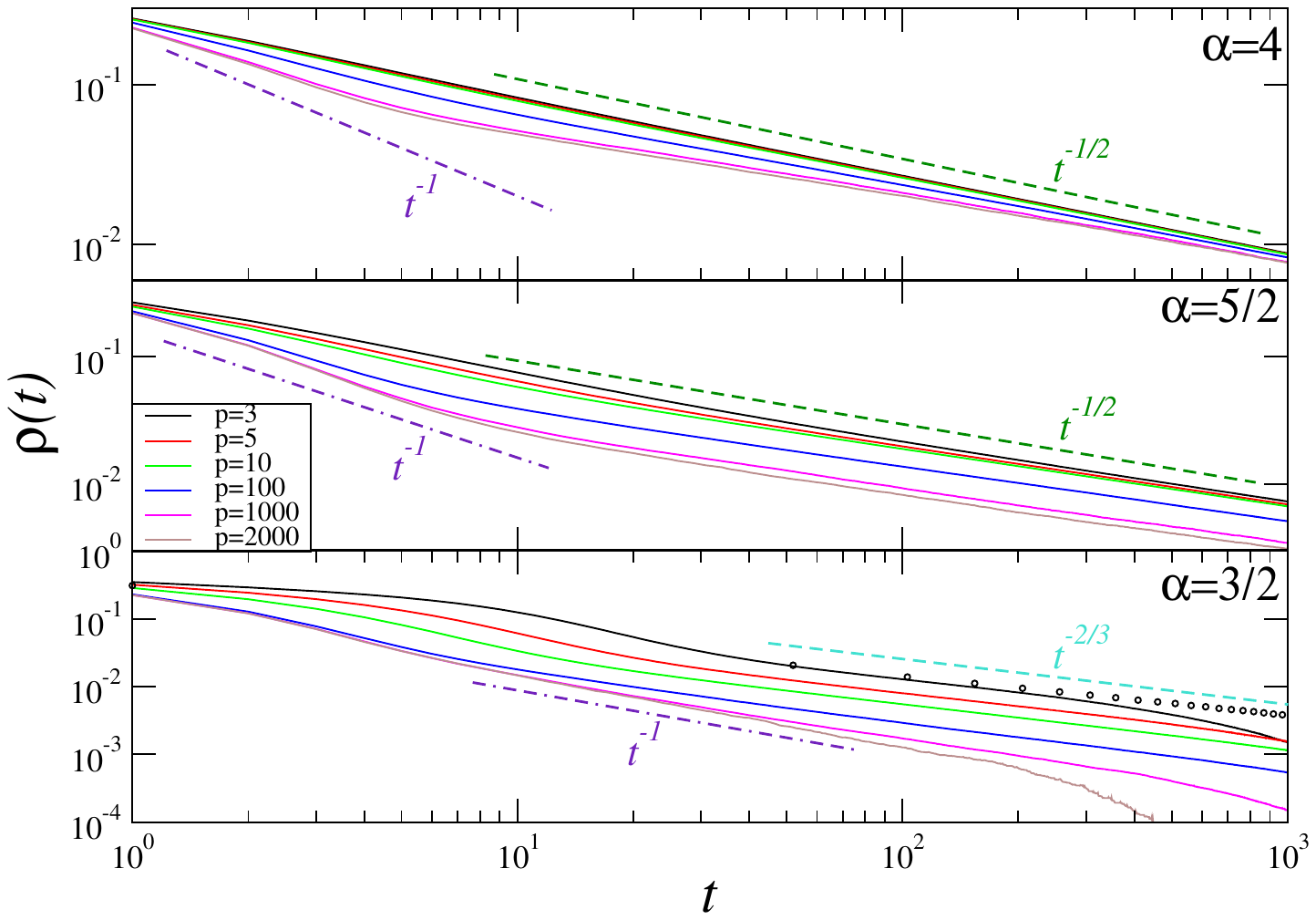}}}
	\caption{Time evolution of $\rho$, in a log-log plot, for $\alpha=4$ (top panel), $\alpha=5/2$ (central panel), and 
	$\alpha=3/2$ (bottom panel). System size is $N=10^3$ (for comparison, only for the case $\alpha=3/2$ with $p=3$, a larger system size $N=10^5$ is plotted with black open circles). A statistical average up to $5\times 10^4$ has been taken. In each panel, different curves refer to various values of $p$ (see legend).
	Dashed lines are the expected slopes $t^{-1/2}$ and $t^{-2/3}$ for the IM (at $T\neq 0$), as reported in table~\ref{table_a}.
The dot-dashed line is the expected slope $t^{-1}$ for the IM quenched to $T= 0$).}
	\label{figAgt1}
\end{figure}

The behavior of $\rho(t)$ is reported in Fig.~\ref{figAgt1}. Each panel corresponds to a value of $\alpha$ representative of the various sectors present in table~\ref{table_a}. Specifically we have chosen 
$\alpha =4$ ($\alpha >\alpha_{SR}$ in both models), $\alpha=5/2$ ($\alpha_{LR}<\alpha \le \alpha _{SR}$ for the VM but
$\alpha >\alpha_{SR}$ for the IM), and $\alpha=3/2$ ($\alpha \le \alpha_{LR}$ for the VM and $\alpha _{LR}\le \alpha \le \alpha_{SR}$ for the IM). This figure shows that, irrespective of the value of $p$, in most cases the asymptotic value of the exponent $a$ matches those of the IM at finite $T$ (see table~\ref{table_a}), namely $a=1/2$ for $\alpha =4$ and $\alpha =5/2$, and $a=1/\alpha=2/3$ for $\alpha=3/2$. These behaviors are represented by dashed lines in Fig.~\ref{figAgt1}). For $\alpha =4$ and $5/2$,
the asymptotic exponent $1/2$ is observed very early for small values of $p$, whereas it is delayed for larger $p$-values due to a pre-asymptotic regime with $a\simeq 1$
that will be further discussed later on. The situation is somewhat more complicated for $\alpha =3/2$. In this case finite-size effects start to be observed at late times, resulting in a downward bending of the curves. 
This is particularly evident for $p=3$
and for $p=1000, 2000$. 
Clearly, performing simulations with a larger system size 
would avoid this effect. This, in fact, is shown in Fig.~\ref{figAgt1} where, for $\alpha =3/2$ (lower panel) the behavior of a larger system of size $N=10^5$ is shown, only for $p=3$, with black open circles. The downward bending of the curve disappears, being replaced by a straight algebraic behavior with exponent $2/3$. However, as we will discuss further below, our scope is to investigate the behavior of the model as $p/N$ is varied up to $p>N$. This would not be possible by using very large $N$, because simulations become too slow for very large values of $p$. 
Indeed, the value $N=10^3$ has been carefully chosen as an optimal compromise between having a sufficiently large size and the possibility to have results with $p>N$ (we arrive up to $p/N=2$).
In addition to finite-size effects (as for $p=3$) the asymptotic exponent $a$ can also be masked by the pre-asymptotic exponent $a=1$ already mentioned. This happens for the two largest values of $p$. However, it is expected that the exponent $2/3$ would be observed if one could go to times when the pre-asymptotic regime is over, without meeting finite-size effects, which of course would require much larger systems.

With these specifications, our findings show that the pVM is
in the Ising universality starting from $p=p^*\equiv 3$. 
Notice that, for $\alpha >2$, the model, resembling the IM, will
eventually equilibrate ($a=0$, last column in table~\ref{table_a}), but this occurs on huge timescales not accessed in the simulations.

The effect of changing $p$ is particularly visible in the pre-asymptotic stage, where a larger $p$ produces a faster decrease of the curves, particularly for 
the smaller values of $\alpha$ until, for sufficiently large values of $p$, a regime with $a=1$ starts to be observed
which widens upon further increase of $p$, pushing the asymptotic behavior to later and later times.
This fact has a profound meaning. Indeed, as discussed at the end of Sec.~\ref{SecModels}, the stochasticity
introduced by the finite value of $p$ become smaller and smaller when this parameter grows and tends to disappear
when $p\to \infty$. In this case, one would expect the pVM to belong to the universality class of the IM 
at $T=0$, because only at zero temperature the IM is deterministic. 
As discussed in Sec.~\ref{SecRecalls}, this class is characterized by $a=1$.
Of course, even if $p$ is large, some
stochastic effect remains and therefore we expect the system to behave asymptotically as the IM at $T\neq 0$,
which we actually see in Fig.~\ref{figAgt1}. However, increasing $p$ the model narrows the case $p=\infty$ and 
a signature of this is left by the $a=1$ behavior observed pre-asymptotically. As expected, this imprint
widens in time as $p$ grows. 

Summarizing, the asymptotic behavior of the pVM falls in the universality class of the IM at small temperatures for any value of $p$, 
the effect of changing this parameter being only the extension of the pre-asymptotic stage where the universality class of the T=0 IM is 
temporarily observed. In view of the discussion contained in Sec.~\ref{SecModels}, one toggles between the two behaviors crossing over
from $p\ll N$ to $p\gtrsim N$. Hence 
it is natural to expect that physical
quantities, specifically $\rho$, should not depend separately on $p$ and $N$, but only on the ratio. Then,
reinstating the $N$ and $p$ dependence in the interfacial density $\rho(t;p,N)$, which we omitted before
for simplicity, we should have
\begin{equation}
	\rho(t;p,N)=R \left(t;\frac{p}{N}\right),
	\label{eqscal}
\end{equation} 
where $R(t;x)$ is a function with
the following limiting behaviors
\begin{equation}
	R(t;x)\propto \left \{ \begin{array}{lr}
		t^{-a_T}, & \hspace{1cm}\mbox{for } x\ll 1, \\
		t^{-a_{0}}, & \mbox{for } x\gtrsim 1 ,
		\end{array} \right .
		\label{eqscal1}
\end{equation}
where $a_0$ and $a_T$ indicate the asymptotic value of the $a$ exponent in the IM quenched to $T=0$ or to a small finite temperature $T$, respectively. Eq.~(\ref{eqscal}) is
meant for large times but before finite-size effects enter the game;
in other words with the order of the limits 
$\lim _{t\to \infty} \lim _{N\to \infty}$. 

\begin{figure}[ht]
	\vspace{1.0cm}
	\centering
	\rotatebox{0}{\resizebox{1.0\textwidth}{!}{\includegraphics{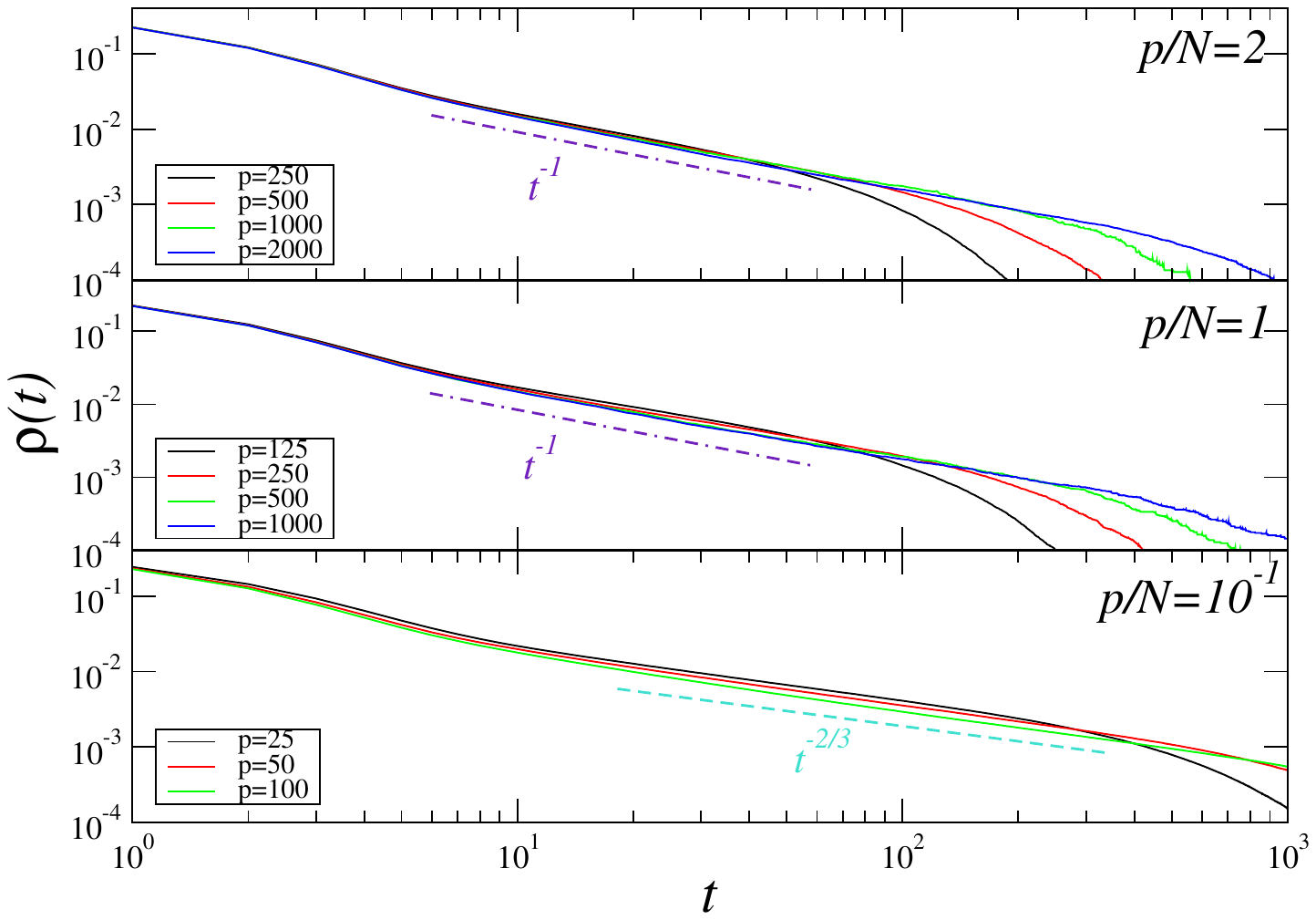}}}
	\caption{Time evolution of $\rho$, in a log-log plot for $\alpha=3/2$ and different values of $p$ (see legend) and system sizes $N$, chosen as to have 
	a fixed value of $x=p/N$ in any case ($p/N=2,1,10^{-1}$ in the upper, central, and lower panel, respectively). A statistical average up to $5\times 10^4$ realizations has been taken. 
	Dashed lines are the expected slope $t^{-1}$ and $t^{-2/3}$ for the IM quenched to $T=0$ or to $T>0$, respectively.}
	\label{fig_mezzi-mezzi}
\end{figure}

We test this conjecture numerically 
in Fig.~\ref{fig_mezzi-mezzi}, for 
$\alpha =3/2$. In each of the three panel we plot $\rho$ for different values of $p$ and system sizes chosen
as to have a fixed (within the same panel) value of $x=p/N$. The three panels correspond to $x=2,1,10^{-1}$,
so to span all the crossover pattern 
described by Eq.~(\ref{eqscal1}). 
Firstly, since at fixed $x$ smaller values of $p$ correspond to smaller system sizes, we see that for smaller $p$ the finite size effects, signaled by the downward late-time bending of the curves, is observed before. 
This being said, looking at sufficiently large times (but before such finite-size effect) one sees that the curves superimpose (for any $x$), according to Eq.~(\ref{eqscal}), and that the decay exponent $a$ matches 
$a_0=1$ for $x\gtrsim 1$ while it 
is consistent with $a_T=2/3$ for 
$x\ll 1$, which is the content of Eq.~(\ref{eqscal1}). 
We found similar results for other values of $\alpha$, with $a$ toggling in any case between 
the corresponding values of $a_0$ and $a_T$. 

\begin{figure}[ht]
	\vspace{1.0cm}
	\centering
	\rotatebox{0}{\resizebox{1.0\textwidth}{!}{\includegraphics{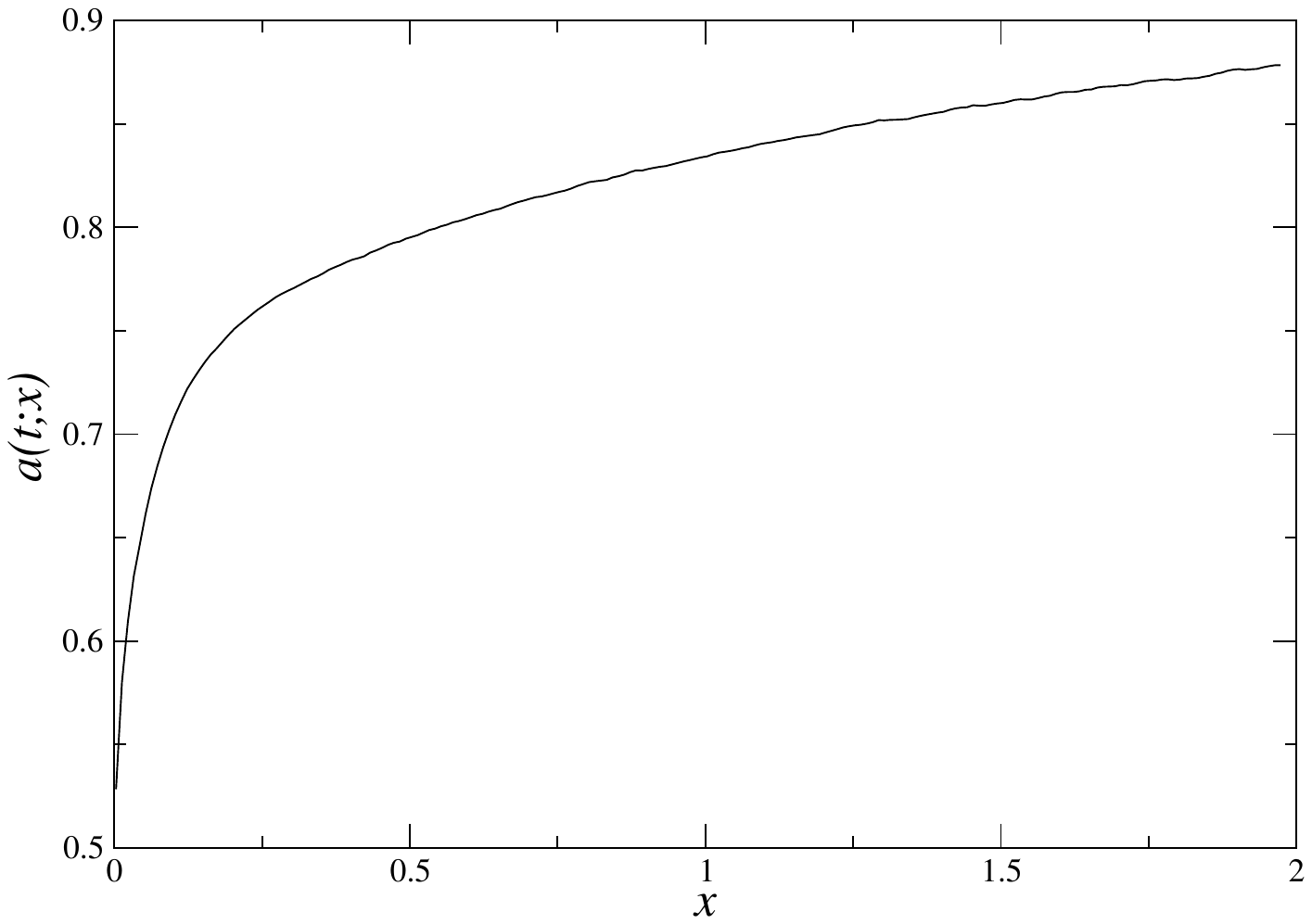}}}
	\caption{The effective exponent $a(t;x)$ defined in Eq.~(\ref{effexp}) is plotted against $x=p/N$, for $\alpha =4$ and $t=3$. System size is $N=10^3$. Statistical average is taken over $2\times 10^4$ realizations.}
	\label{fig_many_p}
\end{figure}

Besides checking data collapse for selected values of $p$ and $N$ in Fig.~\ref{fig_mezzi-mezzi}, we have also computed the whole function 
$R(t;x)$ of Eq.~(\ref{eqscal}), for any value of $x=p/N$. Given the behavior~(\ref{eqscal1}), it is convenient to plot the effective exponent 
\begin{equation}
	a(t;x)=\frac{\partial \ln R(t;x)}{\partial \ln t},
	\label{effexp}
\end{equation} 	
which should toggle between $a_T$ and 
$a_0$ around $x\simeq 1$. This quantity is plotted in Fig.~\ref{fig_many_p} in the case $\alpha =4$, with $t=3$. We see, indeed, that $a(t;x)$ raises from 
$a_T=1/2$ towards $a_0=1$, as expected.

All the previous results were uniquely based on the inspection of $\rho (t)$. However the concept of universality class refers to a whole bunch of universal exponents. For this 
reason in the following,  we repeat a similar analysis computing another quantity, the autocorrelation function
\begin{equation}
	A(t,t')=\langle S_i(t)S_i(t')\rangle,
\end{equation} 
with $t\ge t'$. Notice that this quantity does not depend on $i$ due to the homogeneity of the model. This can be enforced by computing $A(t,t')=\frac{1}{N}\sum _{i=1}^N 
\langle S_i(t)S_i(t')\rangle$ in numerical simulations.
For large time differences $t-t'$ the autocorrelation function decays algebraically as
\begin{equation}
	A(t,t')\propto \left ( \frac{t}{t'}\right )^{-b},
\end{equation}
both in the 1d-voter model~\cite{corberi2024aging} (but only for $\alpha >1$, see discussion below) and in the Ising model~\cite{Corberi_2019,PhysRevE.102.020102}. The value of the exponent $b$ depends on $\alpha$ as illustrated in table~\ref{table_b}.

\begin{table}[ht]
	\centering
	\begin{tabular}{|c| |c|}
		\hline \hline
		\multicolumn{2}{|c|}{\bf{VM}} \\
		\hline \hline
		$0\le \alpha \le 1\, $ &
		\hspace{0.5cm}exponential \hspace{0.5cm} \\
		\hline
		$1< \alpha \le \alpha_{LR}\, (=2)$ & {$1/(\alpha -1)$} \\
		\hline
		\hspace{0.5cm}$\alpha_{LR}<\alpha\le\alpha_{SR}\,(=3)$\hspace{0.5cm} &{$1/(\alpha -1)$}\\
		\hline
		$\alpha >\alpha_{SR}$ & {$1/2$}\\ 
		\hline \hline
		\multicolumn{2}{|c|}{{\bf IM}} \\
		\hline \hline
		$0\le \alpha \le \alpha_{LR}\,(=1)$ &{?} \\
		\hline
		\hspace{0.5cm}$\alpha_{LR}<\alpha\le \alpha_{SR}\,(=2)$\hspace{0.5cm} & {$1/(2\alpha)$} \\
		\hline 
		$\alpha >\alpha_{SR}$  & 1/2 \\
		\hline \hline
	\end{tabular}
	\caption{Value of the exponent $b$ for (top of the table) the VM (after Ref.~\cite{corberi2024aging}) and for (bottom of the table) the IM quenched to a small but finite temperature (after Ref.~\cite{Corberi_2019}), for different ranges of $\alpha$, specified in the left column (values of $\alpha_{LR}$ and $\alpha_{SR}$ are also indicated in parenthesis).  
	The entry $?$ for $0\le \alpha \le \alpha_{LR}$ in the IM indicates that the value of $b$ is not known
	in this case.}
	\label{table_b}
\end{table}
Notice that the value of $b$ is not known for the IM when $\alpha \le \alpha _{LR}$, which is represented by a question mark in table.

\begin{figure}[ht]
	\vspace{1.0cm}
	\centering
	\rotatebox{0}{\resizebox{1.0\textwidth}{!}{\includegraphics{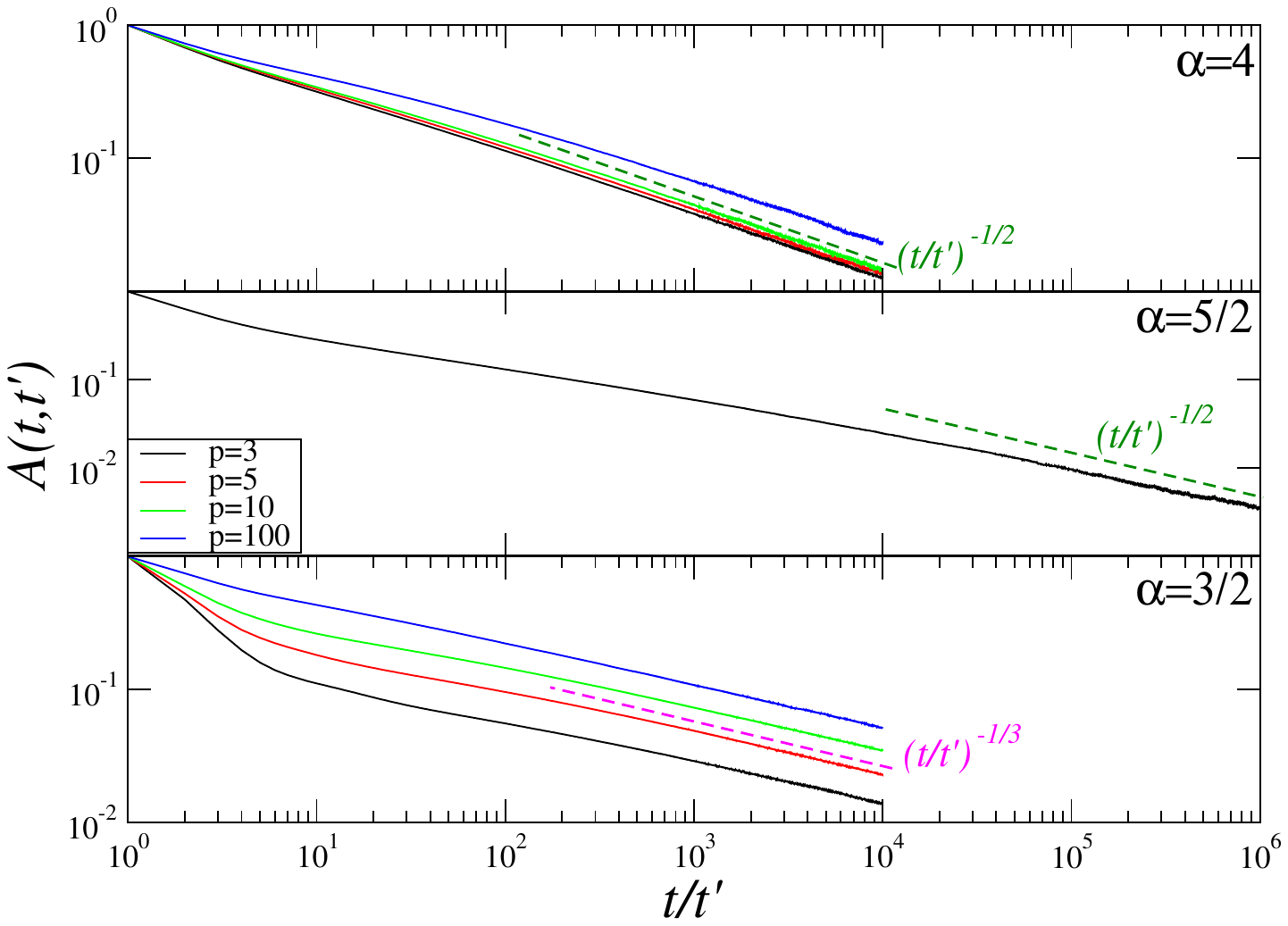}}}
	\caption{The autocorrelation function $A(t,t'=1)$ is plotted in a log-log plot, for $\alpha=4$ (top panel), $\alpha=5/2$ (central panel), and 
	$\alpha=3/2$ (bottom panel). System size is $N=10^5$ for $\alpha=3/2$ and $4$, and $N=10^4$ for $\alpha =5/2$. A statistical average up to 6500 realizations has been performed. In each panel, different curves refer to various values of $p$ (see legend).
	Dashed lines are the expected slopes ($(t/t')^{-1/2}$ and $(t/t')^{-1/3}$ for the IM (at $T\neq 0$), as reported in table~\ref{table_b}.}
	\label{fig_autocorr}
\end{figure}

Our data for the autocorrelation function are shown in Fig.~\ref{fig_autocorr}. Since $b$ is not expected to depend on $t'$ we set the small value $t'=1$.
Simulations for $A(t,t')$ are more time consuming than those for $\rho (t)$. For this reason we could not access values of $p$ as large as $p=1000, 2000$. Furthermore, in the case $\alpha=5/2$ we had to collect data up to times $t-t'=10^6$, because the convergence to the asymptotic value of $b$ is very slow, as it can be appreciated in Fig.~\ref{fig_autocorr} (central panel). Since this is a rather hard numerical task, in this case we have only considered the case $p=3$. 

The data presented in Fig.~\ref{fig_autocorr} show that the observed value of $b$ is very well consistent with the one expected for the IM reported in table~\ref{table_b}.
Therefore the study $b$ leads to the same conclusions already drawn from the behavior of $a$: for any $p\ge 3$ the model falls into the IM universality class. Regarding the crossover to the 
behavior of the IM at $T=0$, which was evidenced before (considering $\rho (t)$), we cannot comment on this with the present data for $A(t,t')$, due to the limited value of $p$ we could access.

\subsection{\boldmath{$\alpha \le 1$}} \label{a<1}

We have already discussed the fact that the IM with $\alpha \le 1$,
exhibits formation and coarsening of domains only in a fraction
$P_{\alpha}$ of its dynamical realizations. The remaining cases are 
characterized by more homogeneous configurations where, after a microscopic time, all the spins take the same sign and $\rho$ vanishes. 
$P_{\alpha }$ decreases
upon lowering $\alpha$ until $P_{\alpha}=0$ for $\alpha=0$, corresponding
to the full mean-field situation. On the other hand, in the voter model a similar
phenomenon is not observed and all the trajectories have, besides the
intrinsic randomness, a qualitatively similar behavior as the metastable
stationary state is approached, with a finite amount of interfaces 
present in the system. This is true until finite-size effects become relevant
and consensus, namely a fully ordered state, is eventually achieved.
The time of this occurrence, however, diverges with $N$ and hence full ordering is impossible in the thermodynamic limit.
The first thing we have to address, therefore, is
whether the pVM conforms to one of these two contrasting behaviors.
Our simulations show that in the pVM interfaces are present at any time
(clearly, still limiting to the time domain where the finite-size effects discussed 
above are absent). We arrived at this conclusion by computing the fraction 
$P_{\alpha}$ at time $t=1$. In the case $\alpha=0.95$, for instance,
we found that domains where always found, i.e. $P_\alpha =1$ over 
the $5\cdot 10^4$ realizations considered.

In~\cite{CORBERI2023113681}, considering the IM with $\alpha \le 1$
quenched to $T\equiv 0$ and pruning only the configurations exhibiting
coarsening, it was conjectured that $a=1$. However such hypothesis relies
on poor numerical evidences, due to huge noise and important finite-size effects,
and on an uncontrolled analytical approximation. On the contrary,
nothing is known about $a$ for quenches to finite temperatures.

\begin{figure}[ht]
	\vspace{1.0cm}
	\centering
	\rotatebox{0}{\resizebox{1.0\textwidth}{!}{\includegraphics{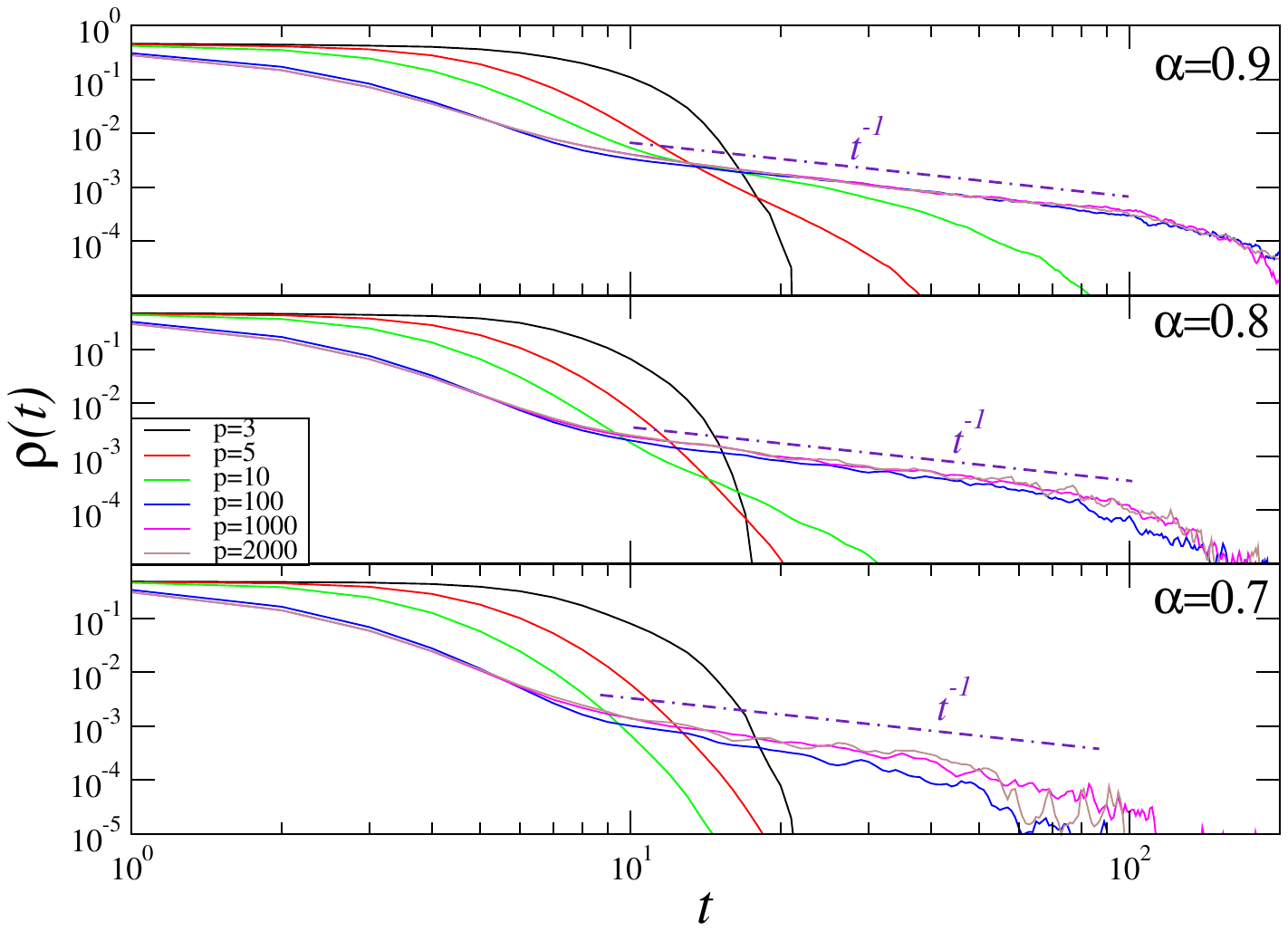}}}
	\caption{Time evolution of $\rho$, in a log-log plot, for $\alpha=0.9$ (top panel), $\alpha=0.8$ (central panel), and 
	$\alpha=0.7$ (bottom panel). In each panel, different curves refer to various values of $p$ (see legend).
	Dashed lines are the slope with $a=1$ conjectured for the IM at $T\equiv 0$), as reported in table~\ref{table_a}.} 
	\label{figAle1}
\end{figure}

Our result for the density of interfaces in the pVM with $\alpha \le 1$ are reported 
in Fig.~\ref{figAle1}. 
Data become progressively noisier as 
$\alpha $ is lowered, because the longer ranged interactions give rise to a larger correlation among the spins, making any kind of self-averaging less effective (the curves with the larger values of $p$ are the results of 
a 6 month cpu time calculation on a standard processor). The same feature is found in simulations of the IM with $\alpha \le 1$~\cite{iannone2021,CORBERI2023113681}.
 
If a picture similar to the cases with $\alpha >1$
still applies now, we should expect to see the putative exponent 
$a=1$ at least for $p\to \infty$, where the pVM should match the
universal behavior of the IM at $T\equiv 0$, namely an exponent $a=1$. This is actually seen in the
figure. Indeed, for $p=1000, 2000$ and sufficiently long times ($t\gtrsim 10$) one sees a regime well consistent with $a=1$, before finite size effects
bend downwards the curves around $t\simeq 10^2$. On the other hand, the behavior of 
$\rho$ is very different when smaller values of $p$ are considered, in that a much faster decay is observed.
In the interpretative scheme put forward insofar, this should correspond to
the behavior of the IM at $0<T<T_{c}$. 
However, since nothing is known about the IM at finite temperatures when $\alpha \le 1$,
we cannot comment further the behavior of these curves.

All the results presented insofar indicate that, for any value of $\alpha$, the
pVM belongs to the IM universality class for any value of $p\ge p^{*}=3$.
An heuristic understanding of that could be the following. According to the discussion 
of Sec.~\ref{SecModels} (around Eq.~(\ref{frequencies})), the convergence of
the pVM to the IM for $p\to \infty$ is due to fading off of fluctuations
in the choice of the $p$ spins influencing the flipping candidate $S_{i}$,
a sort of {\it internal averaging}. Lets stipulate that  $\hat p(N)$ is a certain value
of $p$ such that, for $p\ge \hat p$ the pVM can be regarded sufficiently 
{\it near} to the IM, given a reference tolerance, while it is not so for 
smaller $p$-values. Let us notice that, in order to have an effective averaging by choosing $p$
objects among $N$, it must be $p\sim N$ at least.  Now, lets consider a slightly different model where each 
spin attempts a flip two consecutive times (so it can flip even twice, thus returning
to its original state). In such double move, which can be regarded as the {\it elementary} move of this deformed dynamics, all the other spins are fixed. One can imagine that this different dynamical rule
does not modify to much the long-time behavior.
Upon evolving this model at $p=\hat p/2$ one could argue that the same level
of {\it internal averaging} is obtained as the one present in the original model at
$p=\hat p$, hence its resemblance to the IM would be still satisfactory. Generalizing this idea, we could think to a class of models where
each elementary move is, in reality, made of {$\kappa $} flip attempts. 
This way, the deformed kinetics at $p=\hat p/\kappa$ could be also sufficiently
close to the one of the IM. 

How large $\kappa $ can be chosen? Since we are interested in the behavior
of the system on large timescales, and times are naturally measured in Monte Carlo steps, where $N$ spins flips are attempted, we argue that,
in order not to change the large-time behavior, it must be $\kappa/N \lesssim 1$.
This mean that the smallest value $p^{*}$ that can be reached, under the 
condition of staying close enough to the IM, is of ${\cal O}(1)$. This
heuristic argument shows that $p^{*}$ does not scale with $N$, however clearly it does
not explain the precise value $p^{*}=3$ found in our studies. Let us remark,
however that such value is the smallest one where the non-linearity of $\sigma$
(see Eq.~(\ref{eqSigma})) plays a role.

\section{Conclusions} \label{SecConcl}

In this paper we have considered the ordering kinetics of the pVM with algebraic long-range interactions.  In this system, individual spins $S_i$ flip according to a local field $H_i$ produced by the action of $p$ other agents selected randomly at distance $r$ with a
probability decaying algebraically with $r$ (Eq.~\ref{eqP})).   
This ferromagnetic model
reduces to the usual VM with long-range interactions for $p=1$, while it is expected to be equivalent to the IM for large $p$. For generic $p$ it can be considered as a kind of  interpolation between the two. 
Our study shows that for $p=2$ the
model can be mapped exactly on the usual VM (namely $p=1$), whereas 
for any other value of $p$ its properties, inferred by the asymptotic time-decay of the interface density $\rho $ and of the autocorrelation function $A(t,t')$, fall into the universality class of the 
IM after a quench to a finite $T$.
This is very likely due to the action of the non-linearity, effectively acting only for $p\ge 3$ (see Sec.~\ref{SecModels} and Appendix~\ref{Sec_App}).
Moreover, a behavior in the universality class of the IM quenched to $T=0$ is observed  preasymptotically in a time domain 
which diverges as $p\to \infty$. This is embodied by a scaling form (Eq.~(\ref{eqscal})) 
where $p$ and $N$ enter $\rho $ only through their ratio $x=p/N$.
This can be rationalized by interpreting the random choice of the $p$ spins as a sort of effective temperature which is bound to vanish in the limit $p\to \infty$ due to the perfect averaging obtained in this way.

Our study can be framed into the more general context of the possible reduction of the degrees of freedom of a theory. In the specific case of the models considered here, it offers the possibility to study the IM with long-range interactions by means of the pVM with $p\ge 3$, offering the great numerical advantage of reducing the computational time from $\sim N^2$ to $\sim pN$. Moreover, although for $p\ge 3$ the pVM is not exactly solvable, approximate calculations could possibly be more accessible than in the IM. Besides that, the very issue of degrees of freedom reduction informs many fields of human knowledge. 
For instance, the matter of this article is also related to the ubiquitous use of the 
stochastic gradient descent technique~\cite{AMARI1993185} in state-of-the-art machine-learning and deep-learning models. In these problems, the fundamental task is to find the minima of a {\it cost function} $F$, in a
very high dimensional space of variables. This
can be interpreted in close analogy with thermodynamics, where
$F$ is a free energy and there are Avogadro numbers of degrees of freedom involved. Indeed, minimization of the thermodynamic potential is
actually the driving-force for the phase ordering process, here considered for the IM. In physical problems, one usually describes such process by means of some continuous differential equations or Monte Carlo evolution, based on
a gradient-descent approach, which amounts to say that the thermodynamic
force is proportional to the gradient of $F$ (plus some random fluctuations due to thermal effects). However, in the machine-learning community, the
so-called stochastic gradient-descent method is mostly used. In this algorithm, the gradient of $F$ is not computed with respect to the whole 
number of variables, but on a restricted subset $p$ of them, called the {\it minibatch}. Returning to the models considered in this article, this is perfectly 
analogous to the procedure adopted in passing from the IM to the pVM.
Stochastic gradient-descent is practical, because the evaluation on the full space is often computationally quite hard, and sometimes
even impossible. After this initial motivation, it was later realized that the algorithm with stochastic gradient-descent, besides speeding-up a lot the
computation, often provides a better level of optimization and generalization. 
However, a complete understanding of the reasons for that, and of the differences between the original gradient-descent method and its minibatch version, is far from being attained, despite some recent progresses~\cite{NEURIPS2020_6c81c83c,Mignacco_2021,Mignacco_2022,Angelini23}. The results presented in this article, although
mostly interpreted in a more physical perspective, might trigger a better understanding of the broad problem of simplifying the description considering a  
smaller subset of variables.

Coming back to the pVM considered in this paper, as pointed out at the end of Sec.~\ref{SecSimul}, the difference between the cases with $p=1,2$ and those with larger values of $p$ amounts to the fact that the theory cannot be linearized in the latter cases. This might suggest that the results found in this paper could not be confined to the one-dimensional case considered here, but hold in full generality.
If this is true, our study might open the way to a more efficient study of the ordering kinetics of the IM with long-range interactions that, at variance with the $1d$ case, are scarcer. Let us mention, for instance, the recent observation~\cite{PhysRevE.103.012108,PhysRevE.103.052122} of a new growth exponent in the phase-ordering 
of a two-dimensional system quenched to $T=0$.

As a final remark, let us briefly compare the present work with the well-known topic of weighted networks embedded in Euclidean space~\cite{Kleinberg2000NavigationIA,PhysRevE.62.6270,Parongama2001,PhysRevE.65.056709,PhysRevE.70.036117,PhysRevResearch.5.023129}. In such models, a regular lattice is enriched with shortcuts between distant sites, where sites at distance $r$ are linked with a probability $P(r) \propto r^{-\alpha}$. 
Despite this form of the link probability is the same used in this article to study the $p$-voter model,
the topics covered in the aforementioned literature are quite different from the issue considered here.
A first major difference is
the fact that links between different sites are, in our case, annealed, namely they are randomly chosen in a different way at each elementary move, whereas the studies mentioned above
consider the quenched case, when connections are fixed once forever.
Furthermore, in our model, the value of $p$ is fixed, while in those cases each node can have a different number of links. 
Another difference is that we do not start from a model with nearest-neighbor interactions by adding long-range interactions; in the present model the short-range component is absent and the model is inherently long-range.
 
Regarding the spin dynamics occurring on such networks, the voter and the Ising models on small-world networks have also been studied~\cite{Castellano_2003,PhysRevE.67.035102}. In these cases, the system exhibit similar features to their long-range counterparts on a regular lattice. However, it never happens that the voter model, falls into the universality class of the Ising model, as observed in the present $p$-voter model for $p \geq 3$.

\section*{Acknowledgments}

F.C. acknowledges financial support by MUR PRIN PNRR 2022B3WCFS.

\appendix

\section{The linear pVM} \label{Sec_App}

We consider here the linear pVM, a modified version of the pVM where the
non-linear function $\sigma(H)$ defined 
in Eq.~(\ref{eqSigma}) is made linear,
$\sigma (H)\to H$. In this case
the transition rate reads 
\begin{eqnarray} \nonumber
W(S_i)&=&\frac{1}{2N}\prod _{m=1}^p \,\sum _{r_m=1}^{N/2}P(r_m)\sum _{k_m=[[i\pm r_m]]}[1-S_i H_i] =
\frac{1}{2N}\prod _{m=1}^p \,\sum _{r_m=1}^{N/2}P(r_m)\sum _{k_m=[[i\pm r_m]]}\left[1-\frac{1}{p}S_i\sum _{n=1}^pS_{k_n}\right]=\\
&=&\frac{1}{2N} \left \{ \frac{1}{p}\prod _{m=2}^p \left [\sum _{r_m=1}^{N/2}P(r_m)\sum _{k_m=[[i\pm r_m]]}1\right ]\sum_{r_1=1}^{N/2}
\sum_{k_1=[[i\pm r_1]]}(1-S_iS_{k_1})
+(1\leftrightarrow 2,3,\dots,p) 
\right \},
\end{eqnarray}
where $(1\leftrightarrow 2,3,\dots,p)$ means the first term in curly brackets with the exchange of label 1 with any of the other from 2 to $p$. Using the property
$\sum _{r=1}^{N/2}P(r)\sum _{k=[[i\pm r]]} 1=1$
we arrive at
\be
W(S_i)=\frac{1}{2N}\sum _{r=1}^{N/2}P(r)\sum _{k=[[i\pm r]]}(1-S_iS_k),
\ee
namely the same transition rate of the VM, Eq.~(\ref{transrates}).

\bibliography{voterbib}
\bibliographystyle{apsrev4-2}
\end{document}